\documentclass[aps,nofootinbib,noshowpacs,showkeys,preprintnumbers,amsmath,amssymb]{revtex4}
\usepackage{graphicx}

\allowdisplaybreaks[4]

\begin{document}

\title{Searching for lepton number violating $\Lambda$ baryon decays mediated by GeV-scale Majorana neutrino with LHCb}
\author{Guangshuai Zhang$^1$}

\author{Bo-Qiang Ma$^{1,2,3}$}
\email{mabq@pku.edu.cn}

\affiliation{$^1$
School of Physics and State Key Laboratory of Nuclear Physics and
	Technology, Peking University, Beijing 100871,
	China\\
	$^2$Center for High Energy Physics, Peking University, Beijing 100871, China\\
$^3$Collaborative Innovation Center of Quantum Matter, Beijing, China}

\begin{abstract}
We consider the lepton-number-violating processes in $\Lambda_b$/$\Lambda_c$ decays mediated by on-shell GeV-scale sterile neutrino. We calculate the branching ratio for the following processes: $\Lambda_b^{0} \rightarrow \mathcal{B}^+ \mu^{-}\mu^- \pi^+$, where $\mathcal{B}^+$ is $\Lambda_c^+$ or proton, and $\Lambda_c^{+} \rightarrow \Lambda \mu^{+}\mu^+ \pi^-$ as function of the mass of the sterile neutrino $m_N$ and the heavy-light mixing coefficient of the extended Pontecorvo-Maki-Nakagawa-Sakata~(PMNS) matrix $|U_{\mu N}|^2$. The effect of finite detector size is included in our calculation. After comparing the theoretical effective branching ratio with the expected experimental ability of LHCb, we give the sensitivity upper bounds on $|U_{\mu N}|^2$ on the ($|U_{\mu N}|^2$-$m_N$) plane. These channels give comparable results with the bounds given by different search strategies (e.g., at Belle, NuTeV, DELPHI and BEBC) in the mass region $m_N \simeq$~0.25~GeV-4.5~GeV, and the limits are stronger in the mass range 2~GeV $<m_N<$ 4.5~GeV. 
\end{abstract}
\keywords{Majorana neutrino; $\Lambda_b/\Lambda_c$ decays; sterile neutrino; heavy-light neutrino mixing parameters}

\maketitle

\section{Introduction}

There have been clues for the existence of right-handed heavy neutrinos from both particle physics and cosmology. By introducing right-handed heavy neutrino with Majorana mass, the see-saw mechanism~\cite{Minkowski:1977sc, GellMann:1980vs, Mohapatra:1979ia, Yanagida:1980xy} offers a well-known explanation for the tininess of normal neutrino mass supported by the observation of neutrino oscillation. On the cosmological side, heavy neutrinos can generate the observed baryon asymmetry of the universe through leptogenesis, which, in turn, requires the mass of neutrinos to be larger than a few MeV~\cite{Canetti:2012kh}, and such heavy neutrinos can also serve as natural candidates for dark matter~\cite{Dodelson:1993je,Shi:1998km,Asaka:2005pn,Canetti:2012vf}. 

It is still unclear whether neutrino is a Majorana particle, i.e., its antiparticle is identical to itself, or not. The most appealing way to establish the Majorana nature of a neutrino is to look for the lepton-number-violating process where the conservation of the total number of lepton is violated by 2 units ($\Delta L = 2$). Up to now, neutrinoless double beta ($0\nu \beta\beta$) decay is the most promising $\Delta L = 2$ channel. The nonobservation of the process can set strong limits on the upper bounds of the effective neutrino mass $m_{ee}$ ($m_{ee}$ is defined as $\sum_{i=1,2,3}U_{e i}^2 m_i$, where $m_i$ are the individual neutrino masses and $U_{e i}$ are the $\nu_{i}$-$e$ mixing matrix elements)~\cite{DellOro:2016tmg}. However, the experimental potential of $0\nu \beta\beta$ decay is limited by the accuracy of the theoretical calculation of the nuclear matrix elements, of which the divergence from different approximation methods is still unignorable (for recent reviews on $0\nu \beta\beta$ decay, see Refs.~\cite{DellOro:2016tmg, Dolinski:2019nrj, Engel:2016xgb}).

\begin{figure}[htb] 
\includegraphics[width=8cm]{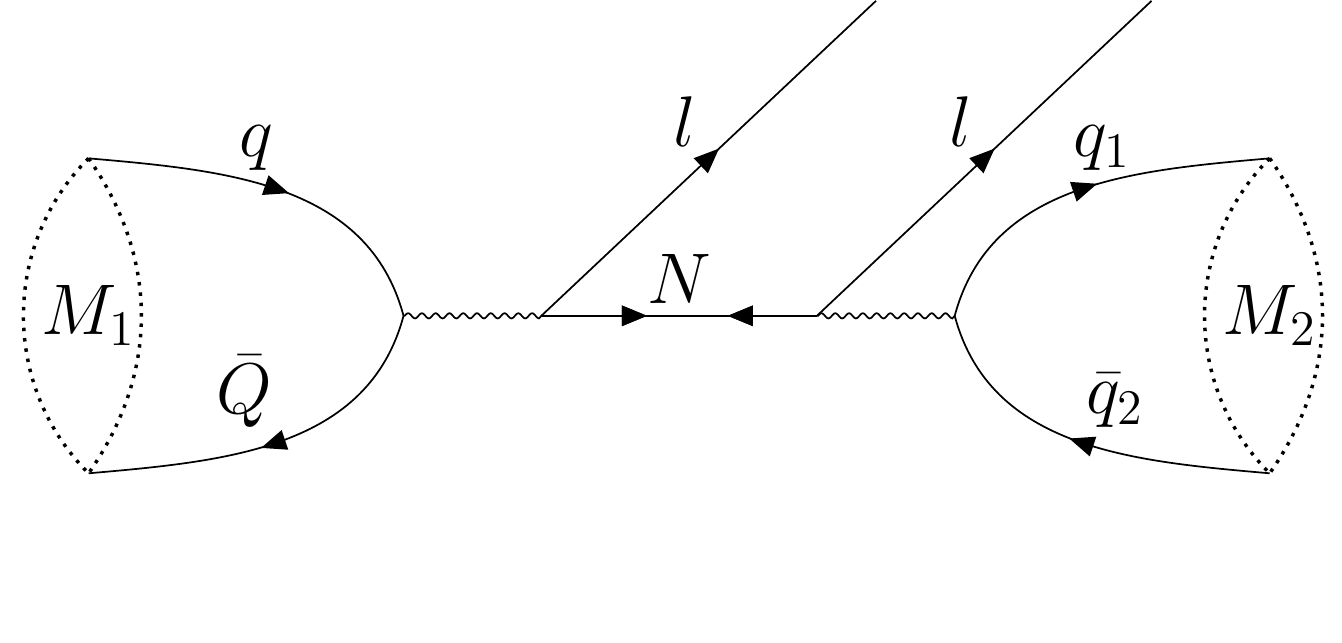} 
\includegraphics[width=8cm]{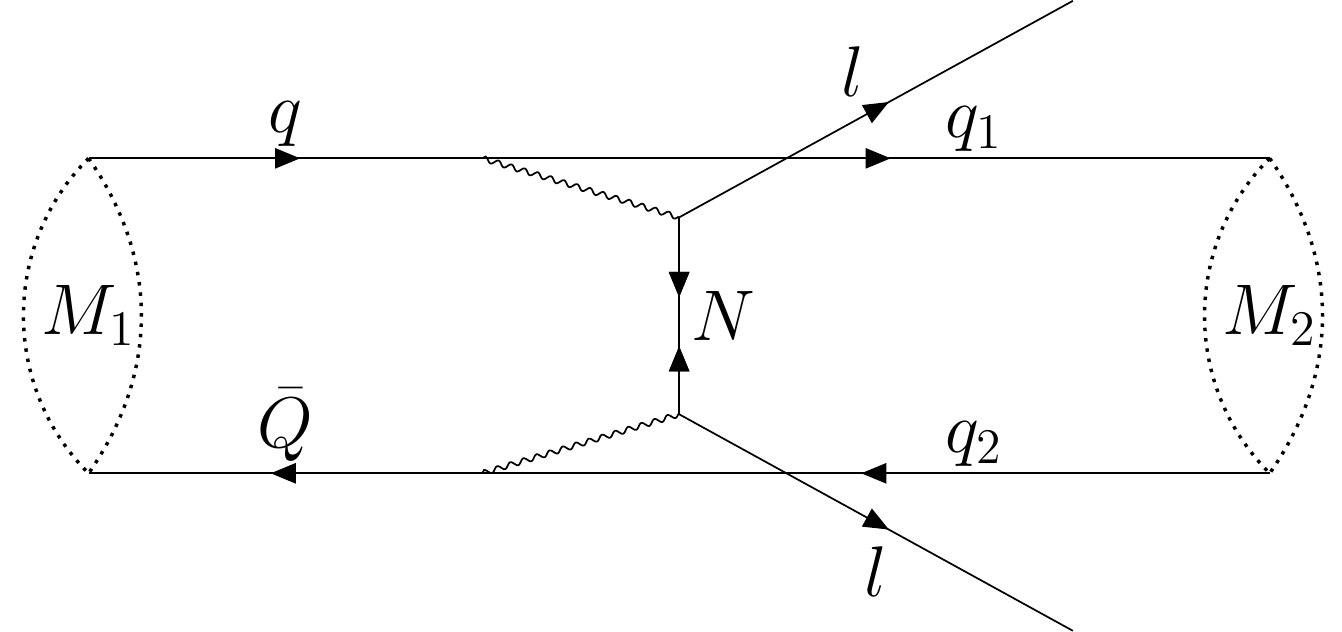}
\caption{ $s$-channel and $t$-channel Feynman diagrams for $M_1 \rightarrow M_2 l l$. $M_1$ is a meson ($K, D, D_s, B, B_c$), $M_2$ is usually $K$ or $\pi$ and $l$'s are leptons. $q$ and $\bar{Q}$ are constituent quarks of $M_1$, while $q_1$ and $\bar{q}_2$ are those of $M_2$. $N$ is the Majorana neutrino.}
\label{M decay}
\end{figure}

As an alternative strategy, lepton-number-violating (LNV) decaying processes in mesons ($K, D, D_s, B, B_c$)~\cite{Abad:1984gh, Littenberg:1991ek, Littenberg:2000fg, Ali:2001gsa, Ivanov:2004ch, Dib:2000wm, Atre:2005eb, Atre:2009rg,Helo:2010cw, Cvetic:2010rw, Cvetic:2016fbv, Cvetic:2017vwl, Cvetic:2019shl,  Milanes:2016rzr, Mejia-Guisao:2017gqp, Milanes:2018aku, Asaka:2016rwd,Chun:2019nwi, Quintero:2011yh}, baryons ($\Sigma^-, \Xi^-, \Lambda_b$)~\cite{Barbero:2002wm, Barbero:2007zm, Barbero:2013fc, Mejia-Guisao:2017nzx}, and $\tau$ lepton~\cite{Castro:2012gi, Gribanov:2001vv, Cvetic:2002jy, Atre:2005eb}, induced by the exchange of the Majorana neutrino (in this paper, we denote the Majorana neutrino with $N$), have been studied extensively in literature. Among these LNV decays, the simplest ones are three-body decays of the form, $M_1 \rightarrow M_2 l l$, where $M_1$ is a meson ($K, D, D_s, B, B_c$), $M_2$ is usually $\pi$ or $K$ meson and $l$'s are leptons. The Feynman diagrams for these processes are shown in Fig.~\ref{M decay}. As is shown by lots of studies, if the neutrino mass is very small ($<$ 1 eV) or very large ($\gg$ 1 GeV), the branching ratios for these LNV decays are very small and they can hardly be observed with current experimental ability~\cite{Abad:1984gh, Littenberg:1991ek, Littenberg:2000fg, Ali:2001gsa, Cvetic:2010rw}. However, Ref.~\cite{Dib:2000wm} shows that, if the mass of the Majorana neutrino lies within the range 245~MeV-389~MeV, i.e., the resonant domain, the branching ratio for the process $K^+ \rightarrow \mu^+\mu^+\pi^-$ can be increased dramatically due to the resonant enhancement of the $s$ channel (the first diagram of Fig.~\ref{M decay}) contribution. The idea was applied to other mesons including $K, D, D_s, B, B_c$ by
Refs.~\cite{Atre:2009rg, Cvetic:2010rw, Helo:2010cw} and $\tau$ lepton by Refs.~\cite{Gribanov:2001vv, Cvetic:2002jy}. The contribution of the $t$ channel, however, is much smaller compared with $s$ channel and its interference with the latter can be overlooked \footnote{Though Ref~\cite{Ivanov:2004ch} shows that, the contribution of $t$ channel may be comparable with that of $s$ channel in a wide range of neutrino mass, it does not affect the magnitude estimate for the decay width while the neutrino mass lies within the resonant domain.}. Thus, narrow width approximation can be applied here due to the on shellness of the Majorana neutrino and the whole process of $M_1 \rightarrow M_2 l l$ can be factorized into two subprocesses: the leptonic decay of $M_1$ into $N l $ and the decay of $N$ into $l M_2$,
\begin{equation}
    \Gamma(M_1 \rightarrow M_2 l l) = \Gamma(M_1 \rightarrow l N)\frac{\Gamma(N \rightarrow  l M_2)}{\Gamma_N},
\label{factorization}
\end{equation}
where $\Gamma_N$ is the total decay width of $N$. As extensions of the idea, the first subprocess on the right of Eq.~(\ref{factorization}) can be replaced with the semileptonic decays of $M_1$, i.e., $M_1 \rightarrow 
 M_3 l N$, where $M_3$ represents another meson, and the second subprocess there can be replaced with leptonic decay of $N$ into $l_1 l_2 \nu_{l_2}$ (Ref.~\cite{Cvetic:2016fbv} gives comprehensive analysis on all possible four-body or five-body LNV decays in B mesons). The resulting processes are four-body or five-body decays, including $B^- \rightarrow D^0/D^* l^- l^- \pi^+$~\cite{Cvetic:2016fbv, Cvetic:2017vwl, Cvetic:2019shl}, $B^- \rightarrow D^0/D^* l^- l^- l_1^+ \nu_{l_1}$~\cite{Cvetic:2016fbv}, $\bar{B}^0 \rightarrow D^+ \pi^+ l^- l^-$~\cite{Quintero:2011yh} $B_c^- \rightarrow J/\psi l^- l^- \pi^+$~\cite{Milanes:2016rzr}, $B_s^0 \rightarrow D_s^-/K^- \pi^- l^+ l^+$~\cite{Mejia-Guisao:2017gqp}, $D_s^+ \rightarrow \phi/\eta \mu^+ \mu^+ \pi^-$~\cite{Cvetic:2019shl} and $D^0 \rightarrow K^+/\pi^+ \pi^+ l^- l^-$~\cite{Milanes:2018aku}. The theoretical branching ratios can be compared to current experimental bounds and stringent limits are set upon the mixing coefficients between sterile neutrino and normal ones in the resonant mass region. Aside from mesons and $\tau$ leptons, LNV processes in baryons, including three-body decays: $\Sigma^- \rightarrow \Sigma^+/p l^- l^-$~\cite{Barbero:2002wm, Barbero:2007zm, Barbero:2013fc}, $\Xi^- \rightarrow \Sigma^+/p l^- l^-$~\cite{Barbero:2007zm}, $\Sigma_c^- \rightarrow \Sigma^+/p l^-l^-$~\cite{Littenberg:1991rd} and four-body decays: $\Lambda_b \rightarrow \Lambda_c^+/p \pi^+ l^- l^-$~\cite{Mejia-Guisao:2017nzx}, are also closely studied in literature.

\begin{figure}[htb]
    \includegraphics[width = 0.4\textwidth]{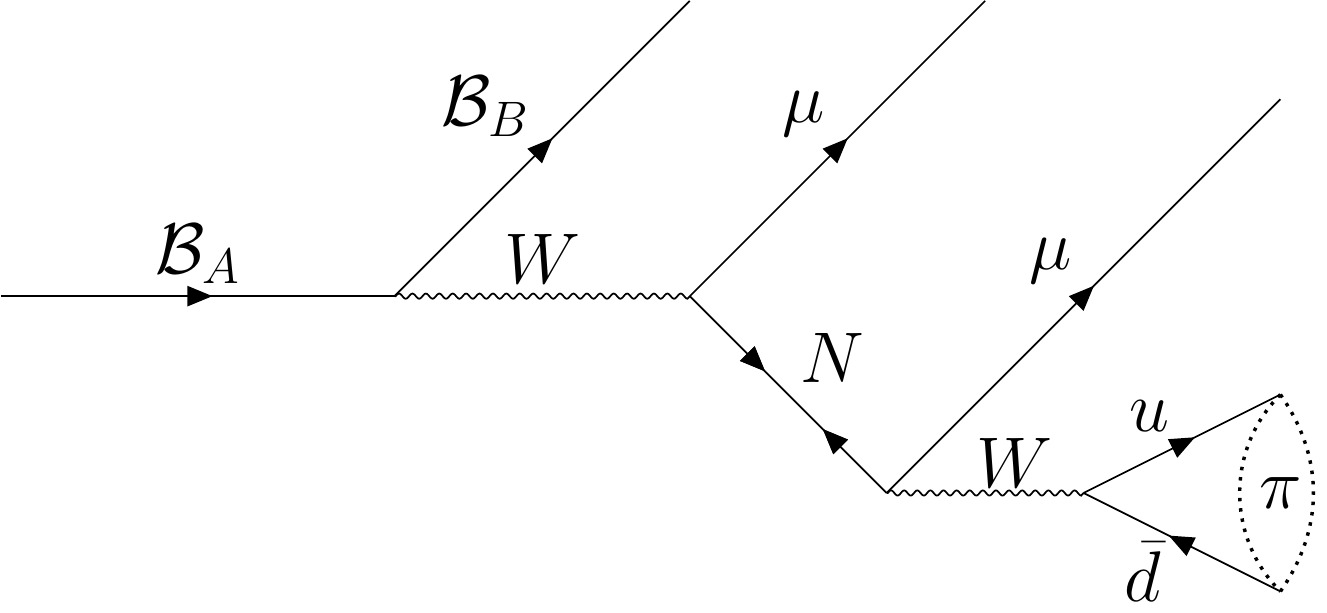}
    \caption{Illustrative Feynman diagram for $\mathcal{B}_A \rightarrow \mathcal{B}_B \pi \mu \mu$. $N$ is the intermediate on-shell Majorana neutrino. If $\mathcal{B}_A$ is $\Lambda_b^0$, then $\mathcal{B}_B$ is $\Lambda_c^+$/proton, the $\mu$'s are double $\mu^-$ and the meson is $\pi^+$, while if $\mathcal{B}_A$ is $\Lambda_c^+$, then $\mathcal{B}_B$ is $\Lambda$, the $\mu$'s are double $\mu^+$ and the meson is $\pi^-$. The crossed diagram with the leptons exchanged must be included.}
    \label{fig:001}
\end{figure}

In this work, we examine the four-body decaying processes of $\Lambda_b$ and $\Lambda_c$: $\Lambda_b^{0} \rightarrow \mathcal{B}^+ \mu^{-}\mu^- \pi^+$, where $\mathcal{B}^+$ is $\Lambda_c^+$ or proton, and $\Lambda_c^{+} \rightarrow \Lambda \mu^{+}\mu^+ \pi^-$ (in this paper we represent the three processes with $\mathcal{B}_A \rightarrow \mathcal{B}_B \pi \mu \mu$). We choose $\Lambda_b$ and $\Lambda_c$ because they are the most produced bottom/charmed baryons in LHCb. The illustrative Feynman diagram for these processes is shown in Fig.~\ref{fig:001}. The analysis is performed in the scenario where there is only one kind of heavy sterile neutrino. The mass for the exchanged neutrino is within the resonant domain, namely $m_{\mu} + m_{\pi} < m_N < m_{\mathcal{B}_A} - m_{\mathcal{B}_B} -m_{\mu}$, so that the resonant enhancement effect dominates. The masses of neutrino and the heavy-light coupling parameters are considered as independent parameters to make a general argument.

This paper is organized as follows. In Section \ref{sec.2}, we give our calculation of the branching ratio for $\mathcal{B}_A \rightarrow \mathcal{B}_B \pi \mu \mu$. Taking account of the resonant enhancement effect, we factorize the whole process into two subprocesses: $\mathcal{B}_A \rightarrow \mathcal{B}_B  \mu N$ and the subsequent decay of on-shell $N$. The effect of the finite detector size is taken into account. In Section \ref{sec.3}, we estimate the expected branching ratio of LHCb and give the numerical results by comparing the theoretical calculation with the experimental ability. Section \ref{sec.4} gives the conclusion.

\section{Calculation of the branching ratio for $\mathcal{B}_A \rightarrow \mathcal{B}_B \pi \mu \mu$}
\label{sec.2}

The lepton-number-violating processes $\mathcal{B}_A \rightarrow \mathcal{B}_B \pi \mu \mu$ can occur via the exchange of an on-shell Majorana neutrino $N$, which means the mass of $N$ is between $m_{\mu} + m_\pi$ and $m_{\mathcal{B}_A} - m_{\mathcal{B}_B} -m_{\mu}$. Thus, the narrow width approximation can be applied here and the total decay width can be factorized as~\cite{Cvetic:2016fbv}
\begin{equation}
\Gamma(\mathcal{B}_A \rightarrow \mathcal{B}_B \pi \mu \mu) = \Gamma(\mathcal{B}_A \rightarrow \mathcal{B}_B  \mu N)\frac{\Gamma(N \rightarrow \mu \pi)}{\Gamma_N},
\end{equation}
where $\Gamma_N$ is the total decay width of $N$. The second factor of the right-hand side is well known (see, e.g., Refs.~\cite{Cvetic:2016fbv,Atre:2009rg}),
\begin{equation}
\Gamma(N \rightarrow \mu \pi) =\frac{G_F^2}{16\pi}|V_{ud}^{\mathrm{CKM}}|^2|U_{\mu N}|^2 f_{\pi}^2 m_N\lambda^{1/2}(m_N^2, m_{\mu}^2, m_{\pi}^2)\left[\left(1 - \frac{m_{\mu}^2}{m_N^2}\right)^2 - \frac{m_{\pi}^2}{m_N^2}\left(1 + \frac{m_{\mu}^2}{m_N^2}\right)\right],
\end{equation}
where $G_F = 1.1664 \times 10^{-5} \mathrm{GeV}^{-2}$ is the Fermi coupling constant, $V_{u d}^{\mathrm{CKM}} = 0.97418$ is the up-down Cabibbo-Kobayashi-Maskawa (CKM) matrix element and $f_\pi$ = 0.1304 GeV is the pion decay constant~\cite{Zyla:2020zbs}. The function $\lambda(x, y, z)$ is the kinematic K{\"a}llen function, $\lambda(x, y, z) = x^2 + y^2 + z^2 - 2xy -2yz- 2xz$. $U_{\mu N}$ is the heavy-light mixing coefficient of the extended Pontecorvo-Maki-Nakagawa-Sakata (PMNS) matrix. For simplicity, we assume that there is only one kind of heavy neutrino, in other words, the light neutrino flavor eigenstate $\nu_l$ (with flavor $l = e, \mu, \tau$) is
\begin{equation}
\nu_{\ell}=\sum_{k=1}^{3} U_{\ell \nu_{k}} \nu_{k}+U_{\ell N} N,
\end{equation}
where $\nu_{k}~(k = 1, 2, 3)$ are the mass eigenstates of light neutrino and $N$ is that of heavy neutrino.

The decay width of the sterile neutrino $\Gamma_N$ can be derived by summing over all of its possible decaying channels including $l^-P^+, ~\nu_l P^0, ~l^-V^+, ~\nu_l V^0, ~l_1^-l_2^+\nu_{l_2}, ~l_1^-l_1^+\nu_{l_2}$ and $\nu_{l_1}\nu \bar{\nu}$ where $P^{+/0}$/$V^{+/0}$ is the pesudoscalar/vector meson, $l ~= ~e,~\mu,~\tau$ and $\nu_l$ is the corresponding neutrino~\cite{Atre:2009rg,Helo:2010cw, Dib:2000wm}. As a result, the decay width is expressed as a function of heavy-light coupling constants and the mass of $N$~\cite{Cvetic:2014nla},
\begin{equation}
\Gamma_N = \tilde{\mathcal{K}} \frac{G_F^2 m_N^5}{96 \pi^3},
\end{equation}
where the $|U_{lN}|$ dependent factor $\tilde{\mathcal{K}}$ is,
\begin{equation}
\tilde{\mathcal{K}} = \mathcal{N}_{e N}\left|U_{e N}\right|^{2}+\mathcal{N}_{\mu N}\left|U_{\mu N}\right|^{2}+\mathcal{N}_{\tau N}\left|U_{\tau N}\right|^{2}.
\end{equation}
The values for the coefficients $\mathcal{N}_{lN}~(l = e,\mu,\tau)$ range from 1 $\sim$ 20 depending on $m_N$. The specific form of $\mathcal{N}_{lN}$ and details for the derivation can be found in Ref.~\cite{Cvetic:2014nla}. In our work, due to its relatively tiny influence on the final result, we set $\mathcal{N}_{lN} \approx 10$ for simplicity, namely,
\begin{equation}
\label{gammaN}
 \Gamma_N \approx 10\sum_{l = e,\mu,\tau}|U_{lN}|^2 \frac{G_F^2 m_N^5}{96 \pi^3}.
\end{equation}

\subsection{Decay width of $\mathcal{B}_A \rightarrow \mathcal{B}_B N \mu $}
\label{2A}

The decay width of the process $\mathcal{B}_A \rightarrow \mathcal{B}_B  \mu N$ is
\begin{equation}
\Gamma(\mathcal{B}_A \rightarrow \mathcal{B}_B  \mu N) = \frac{1}{2m_{\mathcal{B}_A}}\int d\Phi_3 |\mathcal{M}(\mathcal{B}_A \rightarrow \mathcal{B}_B \mu N)|^2,
\end{equation}
where $\mathcal{M}$ is the amplitude for the process and $d\Phi_3$ is the differential three-body phase space,
\begin{equation}
d\Phi_3 = (2\pi)^4 \delta^4(p_{\Lambda_c}- p_{\Lambda} - p_N - p_l)\frac{d^3 \vec{p_{\Lambda}}}{(2\pi)^3 2E_{\Lambda}}\frac{d^3 \vec{p_{N}}}{(2\pi)^3 2E_{N}}\frac{d^3 \vec{p_{\mu}}}{(2\pi)^3 2E_{\mu}}.
\end{equation}
In the $l$-$N$ rest frame~($W^{*}$ frame), it can be simplified as
\begin{equation}
d\Phi_3 = \frac{1}{(2\pi)^5}\frac{1}{16 m_{\mathcal{B}_A}^2}|\vec{p_{\mathcal{B}_B}}||\vec{p_{\mu}^{*}}| d \Omega_{1}^{*} d \Omega_{2}d\sqrt{t},
\label{eq2}
\end{equation}
where $t = (p_N + p_{\mu})^2$, $d\Omega_{1}^* = d\cos\theta_1 d\phi_1$ is the solid angle of $\mu$ in the $\mu$-$N$ frame ($W^*$ frame) and $d\Omega_2 = d\cos\theta_2 d\phi_2$ is that of $\mathcal{B}_B$ in the $\mathcal{B}_A$ rest frame. $\vec{p_{\mathcal{B}_B}}$  is the momentum of $\mathcal{B}_B$ in $\mathcal{B}_A$ frame and $\vec{p_{\mu}^{*}}$ is that of $\mu$ in $\mu$-$N$ frame,
\begin{equation}
|\vec{p_{\mathcal{B}_B}}| = \frac{1}{2m_{\mathcal{B}_A}}\lambda^{1/2}(m_{\mathcal{B}_B}^2, m_{\mathcal{B}_A}^2, t), 
\qquad|\vec{p_{\mu}^*}| = \frac{1}{2\sqrt{t}}\lambda^{1/2}(m_{\mu}^2, m_{N}^2, t).
\label{eq1}
\end{equation}
Putting Eq.~(\ref{eq1}) into Eq.~(\ref{eq2}) and changing the variable from $\sqrt{t}$ to $t$, we write $d\Phi_3$ as
\begin{equation}
d \Phi_3 = \frac{1}{(2\pi)^5}\frac{1}{128 m_{\mathcal{B}_A}t^2} \lambda^{1/2}\left(m_{\mathcal{B}_A}^2, m_{\mathcal{B}_B}^2, t\right)\lambda^{1/2}\left(m_{\mu}^2, m_{N}^2, t\right)d\Omega_1^* d \Omega_2 dt.
\end{equation}

\begin{figure}[htb]
    \includegraphics[width = 0.5\textwidth]{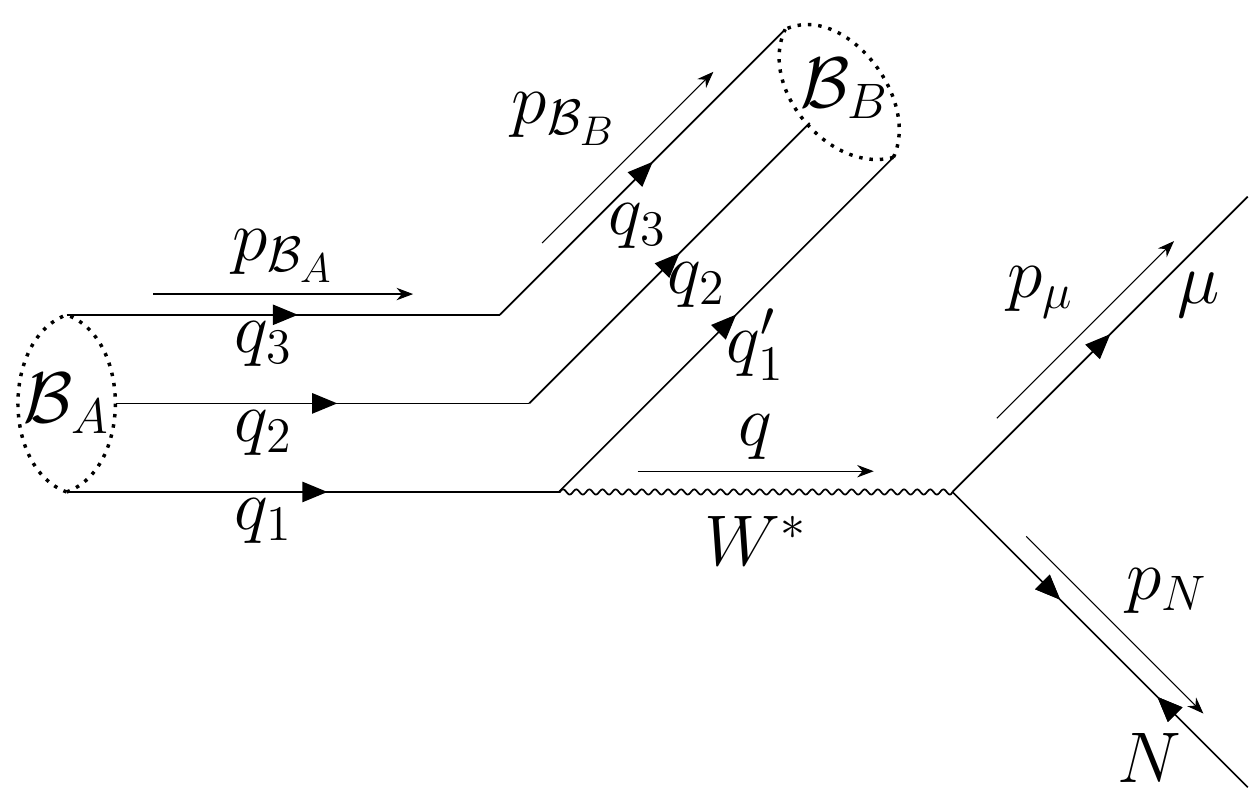}
    \caption{Feynman diagram for $\mathcal{B}_A \rightarrow \mathcal{B}_B N \mu$, where $N$ is the on-shell Majorana neutrino. If $\mathcal{B}_A$ is $\Lambda_b^0$, then $\mathcal{B}_B$ is $\Lambda_c^+$/proton and the $\mu$ lepton is $\mu^-$, while if $\mathcal{B}_A$ is $\Lambda_c^+$, $\mathcal{B}_B$ is $\Lambda$ and the $\mu$ lepton  is $\mu^+$.}
    \label{fig:002}
\end{figure}

The Feynman diagram for the process is shown in Fig.~\ref{fig:002}. The amplitude can be expressed as
\begin{equation}
\mathcal{M}(\mathcal{B}_A \rightarrow \mathcal{B}_B \mu N) = \frac{G_F}{\sqrt{2}}V_{Q_A Q_B}^{\mathrm{CKM}} U_{\mu N}  \langle\mathcal{B}_B(p_{\mathcal{B}_B})|\bar{Q}_B \gamma_{\alpha}(1 - \gamma_5)Q_{A}|\mathcal{B}_A(p_{\mathcal{B}_A})\rangle\left[\bar{u}(p_{\mu})\gamma_{\alpha}(1 - \gamma_5)v(p_N)\right],
\end{equation}
where $Q_A$ and $\bar{Q}_B$ are the flavor transformed quarks in the process and $V_{Q_A Q_B}^{\mathrm{CKM}}$ is the corresponding CKM matrix element. The hadronic matrix element for $\mathcal{B}_A \rightarrow \mathcal{B}_B$ can be parametrized with six transition form factors ($f_1^V, f_2^V, f_3^V$) and ($f_1^A, f_2^A, f_3^A$) as~\cite{Detmold:2016pkz}
\begin{equation}
\begin{aligned}
\langle\mathcal{B}_B(p_{\mathcal{B}_B})|\bar{Q}_B\gamma_{\alpha}Q_A|\mathcal{B}_A(p_{\mathcal{B}_A})\rangle = &\bar{u_{\mathcal{B}_B}}(p_{\mathcal{B}_B})[\gamma_{\alpha}f_1^V(t) - i\sigma_{\alpha\beta}q^{\beta}\frac{f_2^V(t)}{m_{\Lambda_c}}+q_{\alpha}\frac{f_3^V}{m_{\Lambda_c}}]u_{\mathcal{B}_A}(p_{\mathcal{B}_A}),\\
\langle\mathcal{B}_B(p_{\mathcal{B}_B})|\bar{Q}_B\gamma_{\alpha}\gamma_5 Q_A|\mathcal{B}_A(p_{\mathcal{B}_A})\rangle = &\bar{u_{\mathcal{B}_B}}(p_{\mathcal{B}_B})[\gamma_{\alpha}f_1^A(t) - i\sigma_{\alpha\beta}q^{\beta}\frac{f_2^A(t)}{m_{\Lambda_c}}+q_{\alpha}\frac{f_3^A}{m_{\Lambda_c}}]\gamma_5 u_{\mathcal{B}_A}(p_{\mathcal{B}_A}),
\end{aligned}
\end{equation}
where $q = p_{\mathcal{B}_A}-p_{\mathcal{B}_B}$ and $t = q^2$. $\bar{u}_{\mathcal{B}_B}$ and $u_{\mathcal{B}_A}$ are the spinors of the baryons.

After squaring the amplitude and integrating over the solid angle $d\Omega_2$ and $d\phi_1$ (which simply gives a factor of $8\pi^2$), we get the following expression for the decay width,
\begin{equation}
    \begin{aligned}
    &\Gamma(\mathcal{B}_A \rightarrow \mathcal{B}_B  \mu N)\\ =&\frac{G_F^2}{2^{10}\pi^3m_{\mathcal{B}_A}^3}|V_{Q_A Q_B}^{\mathrm{CKM}}|^2|U_{\mu N}|^2\int_{-1}^{1}d\cos\theta_1\int_{(m_{\mu}+m_N)^2}^{(m_{\mathcal{B}_A} - m_{\mathcal{B}_B})^2} d t\\
    &\times \Big[~|f_1^V(t)|^2 c_1^+(t, \theta_1) + |f_2^V(t)|^2 c_2^+(t, \theta_1) + |f_3^V(t)|^2 c_3^+(t, \theta_1)+  f_1^V(t)f_2^V(t)c_{12}^+(t,\theta_1) +  f_1^V(t)f_3^V(t)c_{13}^+(t, \theta_1) \\
    & ~~~+ f_2^V(t)f_3^V(t)c_{23}^+(t, \theta_1) + |f_1^A(t)|^2 c_1^-(t, \theta_1) + |f_2^A(t)|^2 c_2^-(t, \theta_1) + |f_3^A(t)|^2 c_3^-(t, \theta_1) + f_1^A(t)f_2^A(t)c_{12}^-(t, \theta_1)\\
    & ~~~ +  f_1^A(t)f_3^A(t)c_{13}^-(t, \theta_1) + f_2^A(t)f_3^A(t)c_{23}^-(t, \theta_1) + f_1^V(t)f_1^A(t)c_{11}^{\times}(t, \theta_1) + f_1^V(t)f_2^A(t)(t)c_{12}^{\times}(t, \theta_1)\\
    & ~~~ + f_2^V(t)f_1^A(t)c_{21}^{\times}(t, \theta_1) + f_2^V(t)f_2^A(t)c_{22}^{\times}(t, \theta_1) \Big],
    \end{aligned}
\label{decay width}
\end{equation}
where the coefficients of the form factors are
\begin{align*}
    c_1^{\pm}(t, \theta_1) ~= &\frac{64}{t}\Big[2(p_{\mathcal{B}_B}\cdot p_N)(p_{\mathcal{B}_B}\cdot p_{\mu}) + (p_{\mathcal{B}_B}\cdot p_N)(q \cdot p_{\mu}) + (p_{\mathcal{B}_B}\cdot p_{\mu})(q \cdot p_N) \mp m_{\mathcal{B}_A}m_{\mathcal{B}_B}(p_N \cdot p_{\mu})\Big],\\
    c_2^{\pm}(t, \theta_1) ~= &\frac{32}{t m_{\mathcal{B}_A}^2} \bigg\{m_{\mathcal{B}_B}\Big[-2(m_{\mathcal{B}_B} \pm m_{\mathcal{B}_A})(q\cdot p_N)(q\cdot p_{\mu}) + t(m_{\mathcal{B}_B} \mp m_{\mathcal{B}_A})(p_N\cdot p_{\mu})\Big]- 4t(p_{\mathcal{B}_B}\cdot p_N)(p_{\mathcal{B}_B}\cdot p_{\mu})\\
    &+ (q\cdot p_{\mathcal{B}_B})\Big[4(p_{\mathcal{B}_B}\cdot p_N)(q\cdot p_{\mu}) + 4(q\cdot p_N)(p_{\mathcal{B}_B} \cdot p_{\mu}) + 2(q\cdot p_N)(q\cdot p_{\mu}) + t(p_N\cdot p_{\mu})\Big]
    \bigg\},\\
    c_3^{\pm}(t, \theta_1) ~= &\frac{32}{t m_{\mathcal{B}_B}}\Big[q\cdot p_{\mathcal{B}_B} \pm m_{\mathcal{B}_B}(m_{\mathcal{B}_B} + m_{\mathcal{B}_A})\Big]\times\Big[2(q\cdot p_N)(q\cdot p_{\mu}) - t(p_N \cdot p_{\mu})\Big],\\
    c_{12}^{\pm}(t, \theta_1) = &\frac{-64}{t m_{\mathcal{B}_B}}\bigg\{(q\cdot p_N)\Big[(m_{\mathcal{B}_B} \mp m_{\mathcal{B}_A})(p_{\mathcal{B}_B}\cdot p_{\mu}) + 2 m_{\mathcal{B}_B}(q\cdot p_{\mu})\Big] + (m_{\mathcal{B}_B} \mp m_{\mathcal{B}_A}(p_{\mathcal{B}_B}\cdot p_N)(q \cdot p_{\mu}) \\
    &+ (p_N\cdot p_{\mu})\Big[(q\cdot p_{\mathcal{B}_B})(m_{\mathcal{B}_B} \mp m_{\mathcal{B}_A}) + t m_{\mathcal{B}_B}\Big]\bigg\},\\
    c_{13}^{\pm}(t, \theta_1) = &\frac{64}{t m_{\mathcal{B}_B}}\bigg\{(q\cdot p_N)\Big[(m_{\mathcal{B}_B} \pm m_{\mathcal{B}_A})(p_{\mathcal{B}_B}\cdot p_{\mu}) + 2 m_{\mathcal{B}_B}(q\cdot p_{\mu})\Big] + (m_{\mathcal{B}_B} + m_{\mathcal{B}_A}(p_{\mathcal{B}_B}\cdot p_N)(q \cdot p_{\mu}) \\
    &+ (p_N\cdot p_{\mu})\Big[-(q\cdot p_{\mathcal{B}_B})(m_{\mathcal{B}_B} \pm m_{\mathcal{B}_A}) - t m_{\mathcal{B}_B}\Big]\bigg\},\\
    c_{23}^{\pm}(t, \theta_1) = & \frac{64}{t m_{\mathcal{B}_B}}\Big[-2(q\cdot p_{\mathcal{B}_B})(q\cdot p_{\mu}) + t(p_{\mathcal{B}_B}\cdot p_N)(q \cdot p_{\mu}) + t(p_{\mathcal{B}_B}\cdot p_{\mu})(q \cdot p_N)\Big],\\
    c_{11}^{\times}(t, \theta_1) = &\frac{-128}{t}\Big[(p_{\mathcal{B}_B} \cdot p_N )(q \cdot p_{\mu}) - (p_{\mathcal{B}_B} \cdot p_{\mu})(q \cdot p_N)\Big],\\
    c_{12}^{\times}(t, \theta_1) = &\frac{128}{m_{\mathcal{B}_B}}(m_{\mathcal{B}_A} - m_{\mathcal{B}_B})\Big[(p_{\mathcal{B}_B} \cdot p_N )(q \cdot p_{\mu}) - (p_{\mathcal{B}_B} \cdot p_{\mu})(q \cdot p_N)\Big],\\
    c_{21}^{\times}(t, \theta_1) = &\frac{-128}{m_{\mathcal{B}_B}}(m_{\mathcal{B}_A} + m_{\mathcal{B}_B})\Big[(p_{\mathcal{B}_B} \cdot p_N )(q \cdot p_{\mu}) - (p_{\mathcal{B}_B} \cdot p_{\mu})(q \cdot p_N)\Big], \\
    c_{22}^{\times}(t, \theta_1) = &\frac{128}{t m_{\mathcal{B}_B}^2}\Big[2(q \cdot p_{\mathcal{B}_B}) + t \Big]\Big[(p_{\mathcal{B}_B} \cdot p_N )(q \cdot p_{\mu}) - (p_{\mathcal{B}_B} \cdot p_{\mu})(q \cdot p_N)\Big].
\end{align*}
Terms depending on  angle $\theta_1$ here, namely $p_{\mathcal{B}_B}\cdot p_N$ and $p_{\mathcal{B}_B}\cdot p_{\mu}$, are calculated in the $W^*$ rest frame, whereas other terms can be directly written as functions of $t$,
\begin{equation}
\begin{aligned}
    &q\cdot p_{\mathcal{B}_B} = \frac{1}{2}(m_{\mathcal{B}_A}^2 - m_{\mathcal{B}_B}^2 - t), ~q\cdot p_N = \frac{1}{2}(t + m_N^2 - m_{\mu}^2),~ p_{\mathcal{B}_B}\cdot p_N = E_{\mathcal{B}_B}E_N + |\vec{p}_{\mathcal{B}_B}||\vec{p}_N|\cos\theta_1, \\
    &q\cdot p_{\mu} = \frac{1}{2}(t - m_N^2 + m_{\mu}^2),\qquad p_N \cdot p_{\mu} = \frac{1}{2}(t - m_N^2 - m_{\mu}^2),~ p_{\mathcal{B}_B}\cdot p_{\mu} = E_{\mathcal{B}_B}E_{\mu}-|\vec{p}_{\mathcal{B}_B}||\vec{p}_{\mu}|\cos\theta_1,
\end{aligned}
\end{equation}
where $E_{\mathcal{B}_B}$, $E_{\mu}$ and $E_N$ are the energy of $\mathcal{B}_B$, $\mu$ and $N$ in the $W^*$ rest frame, and $\vec{p}_{\mathcal{B}_B}$, $\vec{p}_N$ are the corresponding momenta,
\begin{equation}
\begin{aligned}
    &E_{\mathcal{B}_B} = \frac{1}{2\sqrt{t}}(m_{\mathcal{B}_A}^2 - m_{\mathcal{B}_B}^2 - t),~E_N = \frac{1}{2\sqrt{t}}(t + m_N^2 - m_{\mu}^2),~E_{\mu} = \frac{1}{2\sqrt{t}}(t - m_N^2 + m_{\mu}^2),\\
    &|\vec{p}_{\mathcal{B}_B}| = \frac{1}{2\sqrt{t}}\lambda^{\frac{1}{2}}(m_{\mathcal{B}_A}^2, m_{\mathcal{B}_B}^2, t), ~~|\vec{p}_N| = \frac{1}{2\sqrt{t}}\lambda^{\frac{1}{2}}(m_N^2, m_{\mu}^2, t).
\end{aligned}
\end{equation}
The integration over $\theta_1$ is not performed here because, as shown in the following part, the correction due to the decaying probability of the on-shell neutrino in the detector also depends on $\theta_1$. As for the form factors, we use the lattice results for $\Lambda_b$ decaying channels from Ref.~\cite{Detmold:2015aaa} and for $\Lambda_c$ decaying channels from Ref.~\cite{Meinel:2016dqj}.

\subsection{Decaying probability of the on-shell neutrino in the detector}

In our analysis, the intermediate neutrino is close to its mass shell. Before decaying into $\mu$ and $\pi$, it propagates as a real particle for certain distance, i.e., decaying length. Since the detector only has a finite size, it is possible that only a fraction of the produced neutrinos $N$ decay inside the detector. On such an occasion, only a part of the decaying products of the neutrinos can be detected and the observed total branching ratio is suppressed. This effect has been discussed by a number of authors (see, e.g., Refs.~\cite{Cvetic:2016fbv, Chun:2019nwi, Bonivento:2013jag, Dib:2014iga,  Cvetic:2017vwl, Asaka:2016rwd}). The effective branching ratio is the original one multiplied by the decaying probability of the neutrino inside the detector,
\begin{equation}
    \mathrm{Br_{eff}} = P_N \times \mathrm{Br},
\end{equation}
where 
\begin{equation}
    P_N = 1 - \exp\Big[-\frac{L}{\tau_N \gamma_N \beta_N}\Big].
\end{equation}
Here $L$ is the length of the detector, $\tau_N$ is the lifetime of $N$, $\beta_N$ is the its speed which is usually 1 and $\gamma_N = (1 - \beta_N^2)^{-1/2}$ is the Lorentz time dilation factor. In order to give an estimate of the magnitude of the value, we assume that L = 1~m and $\gamma_N \beta_N \sim 1$. As for $\tau_N = \hbar/\Gamma_N$, we need an estimate of the  magnitude of $|U_{lN}|^2$. Quite a lot of experiments have excluded the region above $10^{-6}$ at $m_N = 1$~GeV (see, e.g., Refs.~\cite{Vaitaitis:1999wq, CooperSarkar:1985nh}). Thus, $\tau_N$ is longer than $1.44 \times 10^{-7}$~s which corresponds to a decay length longer than 52~m, so the decaying probability is smaller than 0.02. This value can be even smaller at a lighter neutrino mass and smaller heavy-light coupling, so its influence on the final result is not negligible.

Taking this into account, in order to give a comprehensive calculation of the effective branching ratio, we need to know the exact form of the decaying probability, or simply, the Lorentz factor $\gamma_N\beta_N = \sqrt{(E_N/m_N)^2 - 1}$ of $N$ in the lab frame. The energy of the on-shell neutrino depends on the energy of mother particle $\Lambda_b$/$\Lambda_c$, the transferred momentum square $t$ as well as the solid angles $d\Omega_1^*$ and $d\Omega_2$. We use the result of Ref.~\cite{Cvetic:2017vwl}, where $E_N$ is expressed as function of kinematic variables of the decaying process,
\begin{equation}
\begin{aligned}
E_N(t;\theta_2;\theta_1,\phi_1)= &\gamma_{\mathcal{B}_A}\{\gamma_W(t)(E_N(t) - \beta_W(t)|\vec{p}_N(t)|\cos\theta_1)\\
&\beta_{\mathcal{B}_A}[\gamma_W(t)(-|\vec{p}_N(t)|\cos\theta_1 + \beta_W(t)E_N(t))\cos\theta_2 - |\vec{p}_N(t)|\sin\theta_1 \cos\phi_1 \sin \theta_2]\}.
\end{aligned}
\end{equation}
Here, $\beta_{\mathcal{B}_A}$ and $\gamma_{\mathcal{B}_A}$ are Lorentz factors of the mother particle $\mathcal{B}_A$ in the Lab frame depending on the energy of the produced $\Lambda_b$/$\Lambda_c$. $|\vec{p}_N(t)| = (\sqrt{t}/2)\lambda^{1/2}(1, m_N^2/t, m_{\mu}^2/t)$ and $E_N = (t + m_N^2 - m_l^2)/(2\sqrt{t})$ are the three-momentum and energy of $N$ in the $\mathrm{W^*}$ rest frame respectively. The meaning of the angles $\theta_1, \phi_1, \theta_2$ can be found in Sec.~\ref{2A}. $\beta_W$ and $\gamma_W$ are Lorentz factors between $W$ boson and $\mathcal{B}_A$, namely,
\begin{equation}
    \gamma_W(t) = \left(1 + \frac{|\vec{q^{\prime}}|^2}{t}\right)^{1/2}, ~ ~ ~ \beta_W(t) = \left(\frac{t}{|\vec{q^{\prime}}|^2} + 1\right)^{-1/2},
\end{equation}
where $|\vec{q^{\prime}}|=(m_{\mathcal{B}_A}/2) \mathrm{\lambda}^{1/2}(1, m_{\mathcal{B}_B}^2/m_{\mathcal{B}_A}^2, t/m_{\mathcal{B}_A}^2)$ is the three-momentum of $\mathrm{W^*}$ in the $\mathcal{B}_A$ rest frame. The details for the derivation of this expression can be found in the Appendix B in Ref.~\cite{Cvetic:2017vwl}. We choose pseudorapidty $\eta = 3$ and transverse momentum $p_T = 10$~GeV for the $\Lambda_b/\Lambda_c$ produced within the acceptance of LHCb and the corresponding energy for the mother particles is $\sim$ 100~GeV. It can be checked that the influence of the energy deviation from this value within the LHCb acceptance (2 $< \eta <$ 5 and 4~GeV $< p_T <$ 25~GeV~\cite{Aaij:2019pqz}) on the final result is negligible. 

As a result, the total effective branching ratio is written as the integration over $t$, $\theta_1$, $\phi_1$, and $\theta_2$~\cite{Cvetic:2017vwl},
\begin{equation}
\begin{aligned}
    \mathrm{Br_{eff}} = &\int_{0}^{2\pi}d\phi_1 \int_{-1}^{1}d \cos \theta_1 \int_{-1}^{1} d\cos \theta_2 \int_{(m_{\mu}+m_N)^2}^{(m_{\mathcal{B}_A} - m_{\mathcal{B}_B})^2} d t\frac{d \Gamma(\mathcal{B}_A \rightarrow \mathcal{B}_B  \mu N)}{d \cos \theta_1 d t}\frac{\Gamma(N \rightarrow \mu \pi)(t)}{\Gamma_{\mathcal{B}_A} \Gamma_N(t)}\\
    &\times P_N(t,\theta_1, \phi_1, \theta_2).
\end{aligned}
\label{Br eff}
\end{equation}

\section{Sensitivity upper bounds on $|U_{\mu N}|^2$}
\label{sec.3}

The expected number of $\Lambda_b$ produced in the next LHCb upgrade is calculated as $\mathcal{L}\times \sigma_{b\bar{b}} \times f(b \rightarrow \Lambda_b)$, where $\mathcal{L} \simeq 40~ \mathrm{fb}^{-1}$ is the expected luminosity the next LHCb upgrade, $\sigma_{b\bar{b}} \simeq 144~ \mathrm{\mu b}$ is the $b\bar{b}$ cross section within the LHCb covered $\eta$ range ($2 < \eta < 5$) \cite{Aaij:2016avz} and $f(b \rightarrow \Lambda_b) \simeq 0.09 $ is the hardronization factor of b quark to $\Lambda_b$ which is calculated from Ref.~\cite{Aaij:2019pqz}\footnote{Ref.~\cite{Aaij:2019pqz} shows that the ratio between the production fraction of $\Lambda_b$ hardrons and the sum of the fraction of $B^-$ and $\bar{B}^0$ is around 0.259 (averaged between $4~\mathrm{GeV} < p_T < 25~\mathrm{GeV}$ and $2 < \eta < 5$), while that between $\bar{B}^0_s$ and the sum of $B^-$ and $\bar{B}^0$ is 0.122. Since $B^{\pm}$, $B^0/\bar{B}^0$, $\bar{B}^{0}_s/B^0_s$ and $\Lambda_b^{0}/\bar{\Lambda}^0_b$ make the majority of $b\bar{b}$ products (private discussion with Yanxi Zhang), the fraction $f(b\bar{b} \rightarrow \Lambda_b)$ is estimated as $0.259/(1+0.259+0.122)\times0.5 \sim 0.09$.}. So, the expected number of $\Lambda_b$ is around $ 40~ \mathrm{fb}^{-1} \times 144~\mathrm{\mu b} \times 0.09 \simeq 5.2
\times 10^{12}$. The number of $D^0$ mesons produced is $\mathcal{L} \times \sigma_{pp \rightarrow D^0}$ where $\sigma_{pp \rightarrow D^0} = 2072~\mathrm{\mu b}$ is the production cross section of $D^0$ from $pp$ collision within the LHCb acceptance~\cite{Aaij:2015bpa}. The ratio between $\Lambda_c$ and $D^0$ meson is around 0.3~\cite{Sun:2019eaa, Kniehl:2020szu}, so the expected number of $\Lambda_c$ is $40~\mathrm{fb}^{-1} \times 2072~\mathrm{\mu b} \times 0.3 \simeq 2.5\times 10^{13}$.

We do a Monte-Carlo simulation to give an estimation of the total detection efficiency of the three decaying channels at LHCb. We generate 10000 samples of the mother particles according to their $p_T$ and $y$ distribution at LHCb, where $p_T$ and $y$ refer to the transverse momentum and rapidity~\cite{Aaij:2017qml, Aaij:2016jht}\footnote{We use the $p_T$ and $y$ distributions of $B^{\pm}$ and $D^0$, which are close to the distributions of $\Lambda_b$ and $\Lambda_c$, respectively.}. Then, we get the four-momenta of the final state particles with randomly chosen phase space variables and calculate the possibility that these particles are within the detection ability of LHCb, namely $2 < \eta < 5$ and $p_T > 0.5~ \mathrm{GeV}$. We also take into consideration that $\Lambda_c$ is reconstructed from $p K^- \pi^+$, while $\Lambda$ is reconstructed from $p \pi^-$ in their detection. We also include the requirement that $\Lambda_c$ and $\Lambda$ decay before the upstream silicon-strip detector which means the decay length of them is less than 2.5~m~\cite{Aaij:2014jba}. Multiplying the tracking efficiency of LHCb 0.96~\cite{Aaij:2014jba}, the final results for $\Lambda_b \rightarrow p \pi \mu \mu$, $\Lambda_b \rightarrow \Lambda_c \pi \mu \mu$ and $\Lambda_c \rightarrow \Lambda \pi \mu \mu$ are around $21\%$, $0.4\%$, and $0.1\%$, respectively. Since a more detailed analysis on the detection efficiency requires more knowledge about the LHCb detectors, which is out of the range of this paper, for now we assume that the estimation based on kinematics of the processes is a good approximation to the real reconstruction efficiency\footnote{Private discussion with the LHCb member Yanxi Zhang.}. The sensitivity upper bound on the heavy-light mixing $|U_{\mu N}|$ at the 95$\%$ confidence is obtained by requiring that $N_{\mathrm{events}} = 3.09$~\cite{Feldman:1997qc}\footnote{The conclusion is only valid under the situation when there is no background or the background is very small. Since the knowledge about the background of the process should be found in experiments and we lack reliable method for estimating the background theoretically, we have to make the assumption that the background is very small here.}. Thus, assuming that the background is very small ($\sim 1$), the expected branching ratios for $\Lambda_b \rightarrow p \pi \mu \mu$, $\Lambda_b \rightarrow \Lambda_c \pi \mu \mu$ and $\Lambda_c \rightarrow \Lambda \pi \mu \mu$ are $3.09/(21\% \times 5.2 \times 10^{12}$), $3.09/(0.4\% \times 5.2\times 10^{12})$, and $3.09/(0.1 \% \times 2.5 \times 10^{13})$, respectively.

The total decay width of the mother particles is $\Gamma_{\Lambda_b} = 1.383\times 10^{-12}$ GeV and $\Gamma_{\Lambda_c} = 3.291\times 10^{-12}$~GeV~\cite{Zyla:2020zbs}. The length of the detector L is set to be 2.3~m, which is the approximate length of the detector of LHCb~\cite{Cvetic:2019shl}.

By requiring that the effective branching ratio is smaller than the expected value of LHCb, we get the limits on the coupling constant $|U_{\mu N}|^2$ at certain mass of $N$. This is done by solving Eq.~(\ref{Br eff}) numerically with the input of neutrino mass $m_N$. Following the common practice (see e.g., refs.\cite{Helo:2010cw, Milanes:2016rzr}), we assume that the three heavy-light mixing coefficients are of the same order, namely $|U_{e N}| \sim |U_{\mu N}| \sim |U_{\tau N}|$ in the decay width of $N$.

It should be noted that, though the channels $\Lambda_b \rightarrow p/\Lambda_c \pi \mu \mu$ are analyzed in similar ways in Ref.~\cite{Mejia-Guisao:2017nzx}, our work differs from the latter in the following aspects: (i) We include refined calculation on the decay probability of the $N$ in the detector, which turns out to be quite small; (ii) Our work relies on the so-called lepton universality assumption, i.e., $|U_{e N}| \sim |U_{\mu N}| \sim |U_{\tau N}|$. It can be checked that even if we take the more conservative assumption that $|U_{eN}|$ and $|U_{\tau N}|$ are of the size of their current upper limit given by other experiments, the result of our work does not change significantly; (iii) We calculate the experimental sensitivity of LHCb upgrade with a different method and the sensitivity we obtained is 1 to 3 magnitudes higher than the ones used in Ref.~\cite{Mejia-Guisao:2017nzx}.

\begin{figure}[htb] 

\includegraphics[width=8.8cm]{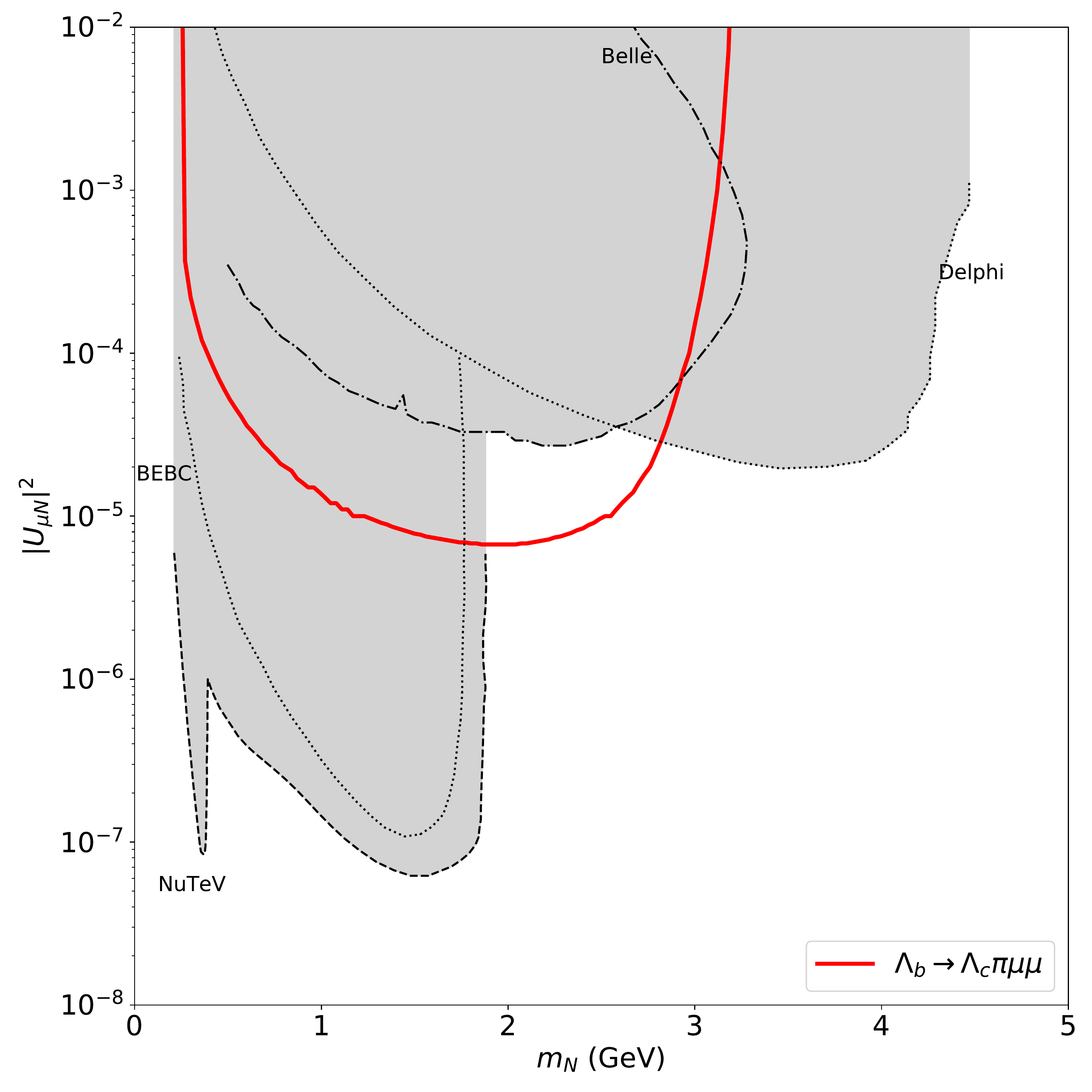} 
\includegraphics[width=8.8cm]{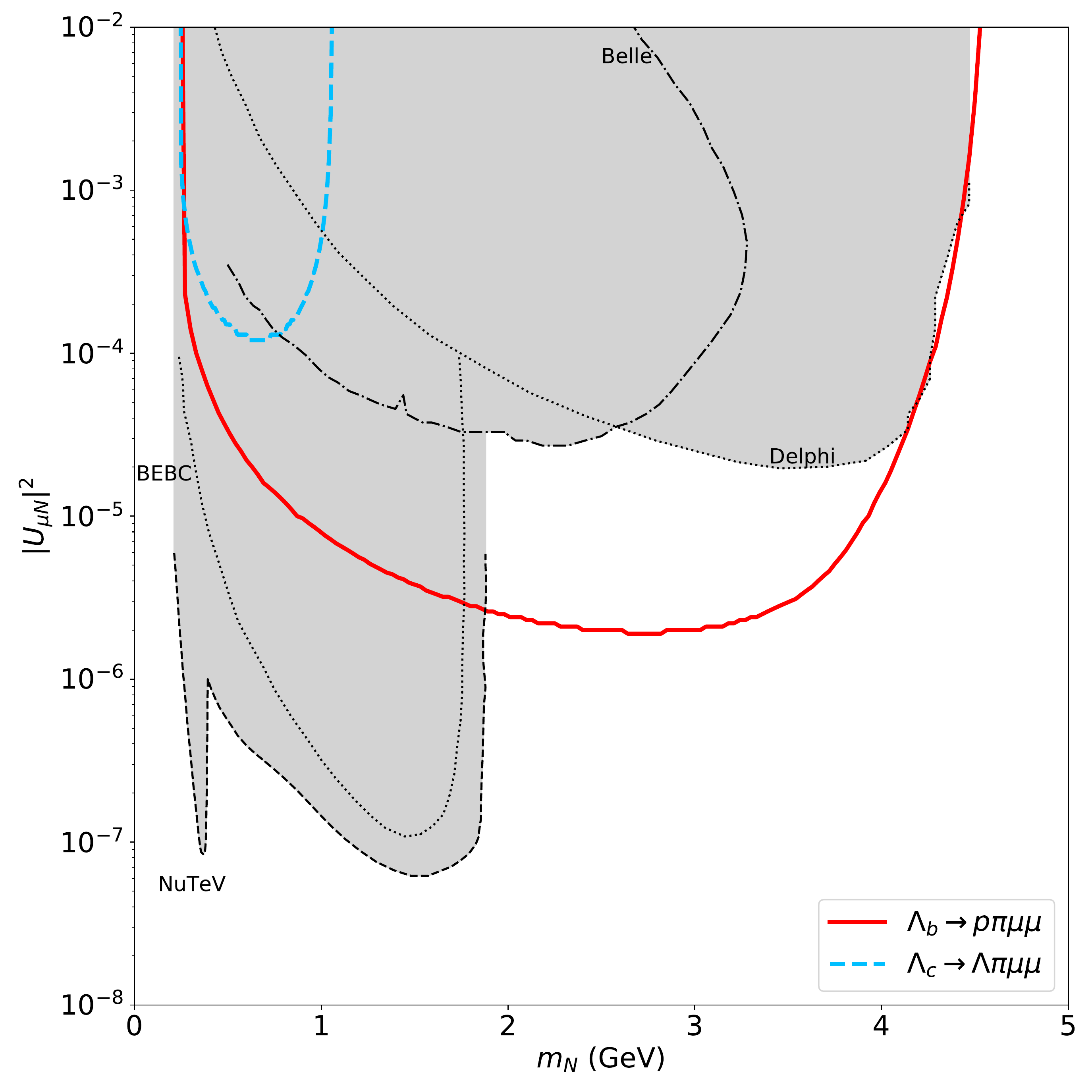}
\caption{Sensitivity upper bounds on $|U_{\mu N}|^2$ by comparing theoretical results with the expected experimental ability of LHCb. The gray areas are excluded by previous experiments, including NuTeV~\cite{Vaitaitis:1999wq}, BEBC~\cite{CooperSarkar:1985nh}, Belle~\cite{Liventsev:2013zz} and Delphi~\cite{Abreu:1996pa}. The colored bold lines are the results we get.}
\label{constraint}
\end{figure}

Our result in shown in Fig.~\ref{constraint}. We compare our results with the constraints given by previous experiments, including NuTeV~\cite{Vaitaitis:1999wq}, BEBC~\cite{CooperSarkar:1985nh}, Belle~\cite{Liventsev:2013zz} and Delphi~\cite{Abreu:1996pa}. It is shown that in the mass region 2~GeV $<m_N<$ 4.5~GeV,  the $\Lambda_b$ decaying channels give stronger constraints with 1 $\sim$ 2 magnitudes lower compared with the current limit, mainly due to the relatively large mass of the $\Lambda_b$ baryon and the low expected branching ratio of LHCb. The constraints given by the $\Lambda_c$ decaying channel, however, do not transcend previous limits, because of the relatively narrow mass range it gives, during which previous experiments have better results. The narrow mass range is mainly due to the small mass difference between $\Lambda_c$ and $\Lambda$.

\section{Conclusion}
\label{sec.4}

In this work, we examine the lepton-number-violating processes in $\Lambda_b$/$\Lambda_c$ decays mediated by on-shell GeV-scale Majorana neutrino: $\Lambda_b^0 \rightarrow \Lambda_c/p \pi^+ \mu^- \mu^-$ and $\Lambda_c^+ \rightarrow \Lambda \pi^- \mu^+ \mu^+$. We do this in a scenario where there is only one kind of sterile neutrino and its mass $m_N$ and heavy-light mixing $|U_{\mu N}|^2$ are independent parameters. The mass of the Majorana neutrino is set between $m_{\mu} + m_{\pi}$ and $m_{\mathcal{B}_A} - m_{\mathcal{B}_B} -m_{\mu}$ so that the resonance enhancement effect is brought in. The branching ratios for these processes are calculated as functions of $m_N$ and $|U_{\mu N}|^2$. The effect of the decaying probability of the intermediate neutrino in the detector is included with a detailed calculation. By comparing the theoretical result with the expected detection ability of the future LHCb upgrade, we give the sensitivity upper bounds on $|U_{\mu N}|^2$ with respect to $m_N$. It turned out that, if these decays were not detected in the future LHCb upgrade, the constraints on the mixing parameters $|U_{\mu N}|^2$ in the mass range 2 GeV $< m_N <$ 4.5~GeV can be significantly improved.

\section{Acknowledgements}
We thank Yanxi Zhang, Yuannin Gao, and Yajun Mao for very helpful advice on the calculation of the expected branching ratio. This work is supported by National Natural Science Foundation of China (Grant No.~12075003).


\begin{thebibliography}{}


\bibitem{Minkowski:1977sc}
P.~Minkowski,
Phys. Lett. B \textbf{67}, 421-428 (1977)
doi:10.1016/0370-2693(77)90435-X

\bibitem{GellMann:1980vs}
M.~Gell-Mann, P.~Ramond and R.~Slansky,
Conf. Proc. C \textbf{790927}, 315-321 (1979)
[arXiv:1306.4669 [hep-th]].

\bibitem{Mohapatra:1979ia}
R.~N.~Mohapatra and G.~Senjanovic,
Phys. Rev. Lett. \textbf{44}, 912 (1980)
doi:10.1103/PhysRevLett.44.912

\bibitem{Yanagida:1980xy}
T.~Yanagida,
Prog. Theor. Phys. \textbf{64}, 1103 (1980)
doi:10.1143/PTP.64.1103

\bibitem{Canetti:2012kh}
L.~Canetti, M.~Drewes, T.~Frossard and M.~Shaposhnikov,
Phys. Rev. D \textbf{87}, 093006 (2013)
doi:10.1103/PhysRevD.87.093006
[arXiv:1208.4607 [hep-ph]].

\bibitem{Dodelson:1993je}
S.~Dodelson and L.~M.~Widrow,
Phys. Rev. Lett. \textbf{72}, 17-20 (1994)
doi:10.1103/PhysRevLett.72.17
[arXiv:hep-ph/9303287 [hep-ph]].

\bibitem{Shi:1998km}
X.~D.~Shi and G.~M.~Fuller,
Phys. Rev. Lett. \textbf{82}, 2832-2835 (1999)
doi:10.1103/PhysRevLett.82.2832
[arXiv:astro-ph/9810076 [astro-ph]].

\bibitem{Asaka:2005pn}
T.~Asaka and M.~Shaposhnikov,
Phys. Lett. B \textbf{620}, 17-26 (2005)
doi:10.1016/j.physletb.2005.06.020
[arXiv:hep-ph/0505013 [hep-ph]].

\bibitem{Canetti:2012vf}
L.~Canetti, M.~Drewes and M.~Shaposhnikov,
Phys. Rev. Lett. \textbf{110}, no.6, 061801 (2013)
doi:10.1103/PhysRevLett.110.061801
[arXiv:1204.3902 [hep-ph]].

\bibitem{DellOro:2016tmg}
S.~Dell'Oro, S.~Marcocci, M.~Viel and F.~Vissani,
Adv. High Energy Phys. \textbf{2016}, 2162659 (2016)
doi:10.1155/2016/2162659
[arXiv:1601.07512 [hep-ph]].

\bibitem{Dolinski:2019nrj}
M.~J.~Dolinski, A.~W.~P.~Poon and W.~Rodejohann,
Ann. Rev. Nucl. Part. Sci. \textbf{69}, 219-251 (2019)
doi:10.1146/annurev-nucl-101918-023407
[arXiv:1902.04097 [nucl-ex]].

\bibitem{Engel:2016xgb}
J.~Engel and J.~Men\'endez,
Rept. Prog. Phys. \textbf{80}, no.4, 046301 (2017)
doi:10.1088/1361-6633/aa5bc5
[arXiv:1610.06548 [nucl-th]].

\bibitem{Abad:1984gh}
J.~Abad, J.~G.~Esteve and A.~F.~Pacheco,
Phys. Rev. D \textbf{30}, 1488 (1984)
doi:10.1103/PhysRevD.30.1488

\bibitem{Littenberg:1991ek}
L.~S.~Littenberg and R.~E.~Shrock,
Phys. Rev. Lett. \textbf{68}, 443-446 (1992)
doi:10.1103/PhysRevLett.68.443

\bibitem{Littenberg:2000fg}
L.~S.~Littenberg and R.~Shrock,
Phys. Lett. B \textbf{491}, 285-290 (2000)
doi:10.1016/S0370-2693(00)01041-8
[arXiv:hep-ph/0005285 [hep-ph]].

\bibitem{Ali:2001gsa}
A.~Ali, A.~V.~Borisov and N.~B.~Zamorin,
Eur. Phys. J. C \textbf{21}, 123-132 (2001)
doi:10.1007/s100520100702
[arXiv:hep-ph/0104123 [hep-ph]].

\bibitem{Ivanov:2004ch}
M.~A.~Ivanov and S.~G.~Kovalenko,
Phys. Rev. D \textbf{71}, 053004 (2005)
doi:10.1103/PhysRevD.71.053004
[arXiv:hep-ph/0412198 [hep-ph]].

\bibitem{Dib:2000wm}
C.~Dib, V.~Gribanov, S.~Kovalenko and I.~Schmidt,
Phys. Lett. B \textbf{493}, 82-87 (2000)
doi:10.1016/S0370-2693(00)01134-5
[arXiv:hep-ph/0006277 [hep-ph]].

\bibitem{Atre:2005eb}
A.~Atre, V.~Barger and T.~Han,
Phys. Rev. D \textbf{71}, 113014 (2005)
doi:10.1103/PhysRevD.71.113014
[arXiv:hep-ph/0502163 [hep-ph]].

\bibitem{Atre:2009rg}
A.~Atre, T.~Han, S.~Pascoli and B.~Zhang,
JHEP \textbf{05}, 030 (2009)
doi:10.1088/1126-6708/2009/05/030
[arXiv:0901.3589 [hep-ph]].

\bibitem{Helo:2010cw}
J.~C.~Helo, S.~Kovalenko and I.~Schmidt,
Nucl. Phys. B \textbf{853}, 80-104 (2011)
doi:10.1016/j.nuclphysb.2011.07.020
[arXiv:1005.1607 [hep-ph]].

\bibitem{Cvetic:2010rw}
G.~Cvetic, C.~Dib, S.~K.~Kang and C.~S.~Kim,
Phys. Rev. D \textbf{82}, 053010 (2010)
doi:10.1103/PhysRevD.82.053010
[arXiv:1005.4282 [hep-ph]].

\bibitem{Cvetic:2016fbv}
G.~Cvetic and C.~S.~Kim,
Phys. Rev. D \textbf{94}, no.5, 053001 (2016)
[erratum: Phys. Rev. D \textbf{95}, no.3, 039901 (2017)]
doi:10.1103/PhysRevD.94.053001
[arXiv:1606.04140 [hep-ph]].

\bibitem{Cvetic:2017vwl}
G.~Cvetic and C.~S.~Kim,
Phys. Rev. D \textbf{96}, no.3, 035025 (2017)
[erratum: Phys. Rev. D \textbf{102}, no.1, 019903 (2020); erratum: Phys. Rev. D \textbf{102}, no.3, 039902 (2020)]
doi:10.1103/PhysRevD.96.035025
[arXiv:1705.09403 [hep-ph]].

\bibitem{Cvetic:2019shl}
G.~Cveti\v{c} and C.~S.~Kim,
Phys. Rev. D \textbf{100}, no.1, 015014 (2019)
doi:10.1103/PhysRevD.100.015014
[arXiv:1904.12858 [hep-ph]].

\bibitem{Milanes:2016rzr}
D.~Milanes, N.~Quintero and C.~E.~Vera,
Phys. Rev. D \textbf{93}, no.9, 094026 (2016)
doi:10.1103/PhysRevD.93.094026
[arXiv:1604.03177 [hep-ph]].

\bibitem{Mejia-Guisao:2017gqp}
J.~Mejia-Guisao, D.~Milan\'es, N.~Quintero and J.~D.~Ruiz-Alvarez,
Phys. Rev. D \textbf{97}, no.7, 075018 (2018)
doi:10.1103/PhysRevD.97.075018
[arXiv:1708.01516 [hep-ph]].

\bibitem{Milanes:2018aku}
D.~Milan\'es and N.~Quintero,
Phys. Rev. D \textbf{98}, no.9, 096004 (2018)
doi:10.1103/PhysRevD.98.096004
[arXiv:1808.06017 [hep-ph]].

\bibitem{Asaka:2016rwd}
T.~Asaka and H.~Ishida,
Phys. Lett. B \textbf{763}, 393-396 (2016)
doi:10.1016/j.physletb.2016.10.070
[arXiv:1609.06113 [hep-ph]].

\bibitem{Chun:2019nwi}
E.~J.~Chun, A.~Das, S.~Mandal, M.~Mitra and N.~Sinha,
Phys. Rev. D \textbf{100}, no.9, 095022 (2019)
doi:10.1103/PhysRevD.100.095022
[arXiv:1908.09562 [hep-ph]].

\bibitem{Quintero:2011yh}
N.~Quintero, G.~Lopez Castro and D.~Delepine,
Phys. Rev. D \textbf{84}, 096011 (2011)
[erratum: Phys. Rev. D \textbf{86}, 079905 (2012)]
doi:10.1103/PhysRevD.84.096011
[arXiv:1108.6009 [hep-ph]].

\bibitem{Barbero:2002wm}
C.~Barbero, G.~Lopez Castro and A.~Mariano,
Phys. Lett. B \textbf{566}, 98-107 (2003)
doi:10.1016/S0370-2693(03)00773-1
[arXiv:nucl-th/0212083 [nucl-th]].

\bibitem{Barbero:2007zm}
C.~Barbero, L.~F.~Li, G.~Lopez Castro and A.~Mariano,
Phys. Rev. D \textbf{76}, 116008 (2007)
doi:10.1103/PhysRevD.76.116008
[arXiv:0709.2431 [hep-ph]].

\bibitem{Barbero:2013fc}
C.~Barbero, L.~F.~Li, G.~L\'opez Castro and A.~Mariano,
Phys. Rev. D \textbf{87}, no.3, 036010 (2013)
doi:10.1103/PhysRevD.87.036010
[arXiv:1301.3448 [hep-ph]].

\bibitem{Mejia-Guisao:2017nzx}
J.~Mejia-Guisao, D.~Milanes, N.~Quintero and J.~D.~Ruiz-Alvarez,
Phys. Rev. D \textbf{96}, no.1, 015039 (2017)
doi:10.1103/PhysRevD.96.015039
[arXiv:1705.10606 [hep-ph]].

\bibitem{Castro:2012gi}
G.~Lopez Castro and N.~Quintero,
Phys. Rev. D \textbf{85}, 076006 (2012)
[erratum: Phys. Rev. D \textbf{86}, 079904 (2012)]
doi:10.1103/PhysRevD.85.076006
[arXiv:1203.0537 [hep-ph]].

\bibitem{Gribanov:2001vv}
V.~Gribanov, S.~Kovalenko and I.~Schmidt,
Nucl. Phys. B \textbf{607}, 355-368 (2001)
doi:10.1016/S0550-3213(01)00169-9
[arXiv:hep-ph/0102155 [hep-ph]].

\bibitem{Cvetic:2002jy}
G.~Cvetic, C.~Dib, C.~S.~Kim and J.~D.~Kim,
Phys. Rev. D \textbf{66}, 034008 (2002)
[erratum: Phys. Rev. D \textbf{68}, 059901 (2003)]
doi:10.1103/PhysRevD.66.034008
[arXiv:hep-ph/0202212 [hep-ph]].

\bibitem{Littenberg:1991rd}
L.~S.~Littenberg and R.~E.~Shrock,
Phys. Rev. D \textbf{46}, 892-894 (1992)
doi:10.1103/PhysRevD.46.R892

\bibitem{Zyla:2020zbs}
P.~A.~Zyla \textit{et al.} [Particle Data Group],
PTEP \textbf{2020}, no.8, 083C01 (2020)
doi:10.1093/ptep/ptaa104

\bibitem{Cvetic:2014nla}
G.~Cveti\v{c}, C.~S.~Kim and J.~Zamora-Sa\'a,
Phys. Rev. D \textbf{89}, no.9, 093012 (2014)
doi:10.1103/PhysRevD.89.093012
[arXiv:1403.2555 [hep-ph]].

\bibitem{Detmold:2016pkz}
W.~Detmold and S.~Meinel,
Phys. Rev. D \textbf{93}, no.7, 074501 (2016)
doi:10.1103/PhysRevD.93.074501
[arXiv:1602.01399 [hep-lat]].

\bibitem{Detmold:2015aaa}
W.~Detmold, C.~Lehner and S.~Meinel,
Phys. Rev. D \textbf{92}, no.3, 034503 (2015)
doi:10.1103/PhysRevD.92.034503
[arXiv:1503.01421 [hep-lat]].

\bibitem{Meinel:2016dqj}
S.~Meinel,
Phys. Rev. Lett. \textbf{118}, no.8, 082001 (2017)
doi:10.1103/PhysRevLett.118.082001
[arXiv:1611.09696 [hep-lat]].

\bibitem{Bonivento:2013jag}
W.~Bonivento, A.~Boyarsky, H.~Dijkstra, U.~Egede, M.~Ferro-Luzzi, B.~Goddard, A.~Golutvin, D.~Gorbunov, R.~Jacobsson and J.~Panman, \textit{et al.}
[arXiv:1310.1762 [hep-ex]].

\bibitem{Dib:2014iga}
C.~Dib and C.~S.~Kim,
Phys. Rev. D \textbf{89}, no.7, 077301 (2014)
doi:10.1103/PhysRevD.89.077301
[arXiv:1403.1985 [hep-ph]].

\bibitem{Vaitaitis:1999wq}
A.~Vaitaitis \textit{et al.} [NuTeV and E815],
Phys. Rev. Lett. \textbf{83}, 4943-4946 (1999)
doi:10.1103/PhysRevLett.83.4943
[arXiv:hep-ex/9908011 [hep-ex]].

\bibitem{CooperSarkar:1985nh}
A.~M.~Cooper-Sarkar \textit{et al.} [WA66],
Phys. Lett. B \textbf{160}, 207-211 (1985)
doi:10.1016/0370-2693(85)91493-5

\bibitem{Aaij:2019pqz}
R.~Aaij \textit{et al.} [LHCb],
Phys. Rev. D \textbf{100}, no.3, 031102 (2019)
doi:10.1103/PhysRevD.100.031102
[arXiv:1902.06794 [hep-ex]].

\bibitem{Aaij:2016avz}
R.~Aaij \textit{et al.} [LHCb],
Phys. Rev. Lett. \textbf{118}, no.5, 052002 (2017)
[erratum: Phys. Rev. Lett. \textbf{119}, no.16, 169901 (2017)]
doi:10.1103/PhysRevLett.118.052002
[arXiv:1612.05140 [hep-ex]].

\bibitem{Aaij:2015bpa}
R.~Aaij \textit{et al.} [LHCb],
JHEP \textbf{03}, 159 (2016)
[erratum: JHEP \textbf{09}, 013 (2016); erratum: JHEP \textbf{05}, 074 (2017)]
doi:10.1007/JHEP03(2016)159
[arXiv:1510.01707 [hep-ex]].

\bibitem{Sun:2019eaa}
J.~Sun [LHCb],
Nucl. Phys. A \textbf{982}, 683-686 (2019)
doi:10.1016/j.nuclphysa.2018.09.062

\bibitem{Kniehl:2020szu}
B.~A.~Kniehl, G.~Kramer, I.~Schienbein and H.~Spiesberger,
Phys. Rev. D \textbf{101}, no.11, 114021 (2020)
doi:10.1103/PhysRevD.101.114021
[arXiv:2004.04213 [hep-ph]].

\bibitem{Aaij:2017qml}
R.~Aaij \textit{et al.} [LHCb],
JHEP \textbf{12}, 026 (2017)
doi:10.1007/JHEP12(2017)026
[arXiv:1710.04921 [hep-ex]].

\bibitem{Aaij:2016jht}
R.~Aaij \textit{et al.} [LHCb],
JHEP \textbf{06}, 147 (2017)
doi:10.1007/JHEP06(2017)147
[arXiv:1610.02230 [hep-ex]].

\bibitem{Aaij:2014jba}
R.~Aaij \textit{et al.} [LHCb],
Int. J. Mod. Phys. A \textbf{30}, no.07, 1530022 (2015)
doi:10.1142/S0217751X15300227
[arXiv:1412.6352 [hep-ex]].

\bibitem{Feldman:1997qc}
G.~J.~Feldman and R.~D.~Cousins,
Phys. Rev. D \textbf{57}, 3873-3889 (1998)
doi:10.1103/PhysRevD.57.3873
[arXiv:physics/9711021 [physics.data-an]].

\bibitem{Liventsev:2013zz}
D.~Liventsev \textit{et al.} [Belle],
Phys. Rev. D \textbf{87}, no.7, 071102 (2013)
[erratum: Phys. Rev. D \textbf{95}, no.9, 099903 (2017)]
doi:10.1103/PhysRevD.87.071102
[arXiv:1301.1105 [hep-ex]].

\bibitem{Abreu:1996pa}
P.~Abreu \textit{et al.} [DELPHI],
Z. Phys. C \textbf{74}, 57-71 (1997)
[erratum: Z. Phys. C \textbf{75}, 580 (1997)]
doi:10.1007/s002880050370


\end{thebibliography}
\end{document}